\documentclass[aps,prl,reprint,groupedaddress,showpacs,showkeys]{revtex4-1}

\usepackage{graphicx}
\usepackage{dcolumn}
\usepackage{bm}
\usepackage{amsmath}         
\usepackage{lineno}

\begin{document}

\preprint{}

\title{Sign switch of Gaussian bending modulus for microemulsions; a self-consistent field analysis exploring scale invariant curvature energies}

\author{Ramanathan Varadharajan}
\email[Author to whom correspondence should be addressed: R. Varadharajan: ]{ramanathan.varadharajan@wur.nl}
\author{Frans A. M. Leermakers}
\email[Dr. F. A. M. Leermakers: ]{frans.leermakers@wur.nl}
\affiliation{Physical Chemistry and Soft Matter, Wageningen University \& Research Center, Stippeneng 4, 6708 WE Wageningen, The Netherlands.}


\date{\today}

\begin{abstract}
Bending rigidities of tensionless balanced liquid-liquid interfaces as occurring in microemulsions are predicted using self-consistent field theory for molecularly inhomogeneous systems. Considering geometries with scale invariant curvature energies gives unambiguous bending rigidities for systems with fixed chemical potentials: The minimal surface Im3m cubic phase is used to find the Gaussian bending rigidity, $\bar{\kappa}$, and a torus with Willmore energy $W=2 \pi^2$ allows for direct evaluation of the mean bending modulus, $\kappa$. Consistent with this, the spherical droplet gives access to $2 \kappa + \bar{\kappa}$. We observe that $\bar{\kappa}$ tends to be negative for strong segregation and positive for weak segregation; a finding which is instrumental for understanding phase transitions from a lamellar to a sponge-like microemulsion. Invariably, $\kappa$ remains positive and increases with increasing strength of segregation. 
\end{abstract}


\maketitle

Interfaces characterized by dense surfactant packings, such as microemulsions \cite{de1982microemulsions,guha2017creating,labrador2017complex} and biological membranes \cite{tanford1980hydrophobic} that are found naturally or manipulated artificially to be in  a state of near-zero tension, have extensive areas. Often such interfaces feature a spontaneous curvature that manifests in spherical or cylindrical (swollen) micelles \cite{schacht1996oil,gompper1995self}. When a system is tensionless and precisely balanced---typical for single component bilayers and expected for the middle-phase microemulsions---the interface's spontaneous curvature vanishes \cite{safran1994statistical} and ultra-low interfacial energies can be achieved \cite{safran1986origin,scriven1976equilibrium}. Here, the elastic moduli, mean ($\kappa$) and Gaussian  ($\bar{\kappa}$) bending rigidities, control the interface fluctuations and topology, respectively. Such systems show a  first-order phase transition from lamellar to sponge-like phases, e.g., upon an increase of the temperature for nonionic systems, and a change of the salinity for ionic systems \cite{safran1986origin,cates1988random,labrador2017complex}. A pre-eminent challenge is to predict, from a molecular model for such interfaces, a means to induce a sign change in the $\bar{\kappa}$ from negative to positive; this signals the loss of stability of the lamellar, $L_\alpha$, oil-surfactant-water ordering in favor of a phase with saddles, $L_3$ or sponge-like. Another long-standing problem is understanding the relation between surfactant chain architecture and corresponding bending rigidities \cite{PhysRevE.57.6825,kurtisovski2007molecular}.

Earlier theoretical methods \cite{szleifer1988curvature,wurger2000bending}, experiments \cite{gradzielski1996experimental,gradzielski1997droplet,safinya1989universality}, and simulations \cite{PhysRevLett.92.116101,thakkar2011verifying} that attempted to link bending rigidities to molecular properties did not provide information on $\bar{\kappa}$; moreover, the results for $\kappa$  were not consistent with each other. Therefore uncertainties prevail and these persist also because internal checks for presented rigidities are rarely provided. As a result, there exists no accepted molecular level theory that convincingly links molecular characteristics to both mechanical parameters of the interfaces ($\kappa $ and $\bar{\kappa}$). Notably, the missing information for  $\bar{\kappa}$ is remarkable as its magnitude and, in particular, its sign are fundamental to the understanding of microemulsions.

The primary obstacle in establishing a molecular model for determining bending rigidities is the requirement of curving the interface at fixed chemical potentials. In this letter, we propose an elegant protocol with internal checks to find these rigidities. We consider interfaces with scale invariant curvature energies and illustrate the protocol for tensionless, balanced liquid-liquid (L/L) interfaces. In line with experimental findings, we report the existence of a sign switch for $\bar{\kappa}$ which triggers a phase transition from planar to sponge-like phases in middle-phase microemulsions. We focus on the role of the interaction parameter which in strong segregation has a large value and for weak segregation a small value; further, we elaborate on the role of the molecular weights of the solvents and that of the amphiphile.  

Experiments, simulations, and calculations \cite{szleifer1988curvature,wurger2000bending,gradzielski1996experimental,gradzielski1997droplet,safinya1989universality,PhysRevLett.92.116101,thakkar2011verifying} reviewed above have major disadvantages and ambiguities because the systems featured too many complications. We examine a tensionless balanced interface which still is highly relevant to middle-phase microemulsion systems wherein oil and water are separated by a surfactant film with extensive areas and often a complex interface topology. Our focus on tensionless (interfacial tension $\gamma = 0$) balanced (spontaneous curvature $J_0=0$) L/L interface avoids the complications of a finite Laplace pressure (i.e. $\Delta P_L=0$) when imposing some interfacial curvature. Such a model is readily implemented in the Scheutjens-Fleer Self-Consistent Field theory (SF-SCF) for molecularly inhomogeneous systems.  We can consider this idealized system in three different geometries with scale-invariant curvature energies. The latter is essential, as it allows for an analysis in the grand canonical ensemble ($\mu, V, T$), which opens a convincing route to estimate the rigidities: (i) A spherically curved droplet with $\Delta P_L=0$ is used to find $2\kappa + \bar{\kappa}$; (ii) A  minimal Im3m surface (by construction has $\Delta P_L=0$) is used to find $\bar{\kappa}$; (iii) A minimal torus interface is used to find $\kappa$ also for conditions that $\Delta P_L=0$. 

We note that the route to obtain rigidities of balanced tensionless L/L interfaces shows similarities but also important differences from the symmetric freely dispersed lipid bilayers \cite{leermakers2013bending}. For bilayers, we could use the Im3m cubic phase and the spherical vesicle to find $\bar{\kappa}$ and $2 \kappa + \bar{\kappa}$, respectively. The cylindrically curved vesicle could be used to obtain $\kappa$ in two ways: (i) As the number of lipids per unit area is found to be a constant (i.e., not a function of the radius $R$ of the cylindrical vesicle), $\kappa$ was found from the excess Helmholtz energy per unit length, $F^{\sigma}_c$, i.e., $\kappa = R F^{\sigma}_c/\pi$; (ii) Realizing that the grand potential of the cylindrical vesicle per unit length $\Omega_c$ is split up equally into bending energy and stretching energy, $\kappa$ is also found from (half) the grand potential density per unit length, i.e. $\kappa = R\Omega_c/(2\pi)$. However, for the tensionless balanced L/L interface, curved in cylindrical geometry with $\Delta P_L=0$, $\kappa$ can neither be computed from the Helmholtz energy per unit length, nor from the grand potential per unit length, as there is neither a conservation of the number of surfactant per unit area nor a conservation of the chemical potentials of the molecules of the system, cf. Figs. (\ref{fgr:1}c) and (\ref{fgr:1}d) shown below. Importantly, in the L/L interface, we do not find a coincidental equal splitting of curvature and tension energies.

Following Helfrich, by expanding the interfacial tension ($\gamma$) in mean curvature ($J =1/{R_1} + 1/{R_2}$) and Gaussian curvature ($K = {1}/{R_1 R_2}$), with $R_1$, $R_2$ being principle radii of curvature as
\begin{equation}
\gamma(J,K) - \gamma(0,0) = - \kappa J_0 J + \frac{1}{2}\kappa J^2 + \bar{\kappa}K,
\label{helfrichsfree}
\end{equation}
we identify $\gamma(J,K)$ as the appropriate characteristic function that carries the bending information for curving the interface at constant chemical potentials \cite{helfrich1973elastic}. This expansion is the starting point for our analysis of interfacial equilibrium properties, as it appears, refer Fig. (\ref{fgr:1}d), for the balanced L/L interface, we find curved interfaces that exist at chemical potentials equal to that of the ground state (tensionless balanced planar interfaces) not only for the surfactant but also for the two solvents. Note, that in this system, both $\gamma(0,0)=0$ and $J_0=0$, Eqn. (\ref{helfrichsfree}) simplifies to $\gamma(J,K)  = \frac{1}{2}\kappa J^2 + \bar{\kappa}K$.

Within SF-SCF framework, extremizing the mean field free energy for a molecularly inhomogeneous system provides both structural and accurate thermodynamic information \cite{scheutjens1979statistical,fleer1993polymers,leermakers2013bending,kik2010molecular}. We have implemented a coarse-grained molecular model in which there are two types of spherically symmetric segments A and B. These segments are used in two solvents, each with length $n$, $A_n$ and $B_n$ forming the two liquid phases $\alpha$ and $\beta$, respectively, and in a diblock copolymer composed of blocks of equal length $N$, $A_NB_N$. This approach requires molecular partition functions, which are evaluated within a lattice considering the molecules as freely jointed chains. Accordingly, segments fit on lattice sites. The lattice sites are organized as homogeneously curved- or planar layers. Driven by the segregation between the segments, an interface develops on which the lattice geometry imposes the curvature. Segment density gradients can only develop in the direction perpendicular to such interface, as a mean field approximation is implemented in lattice layers `parallel' to the interface. In the absence of density gradients, the model is equivalent to the Flory-Huggins theory. There is just one Flory-Huggins interaction parameter ($\chi$) between monomers A and B. We choose a value slightly above the critical point of the binary solvent ($\chi^{\rm cr}=2/n$). Below the minimum value used for $n$ is $4$, and the interaction is chosen between $\chi = 0.52$ - $0.68$ (for more details on method and model refer supplemental material \cite{supplemental}).
\begin{figure}[t]
 \centering
 \includegraphics[width=0.45\textwidth]{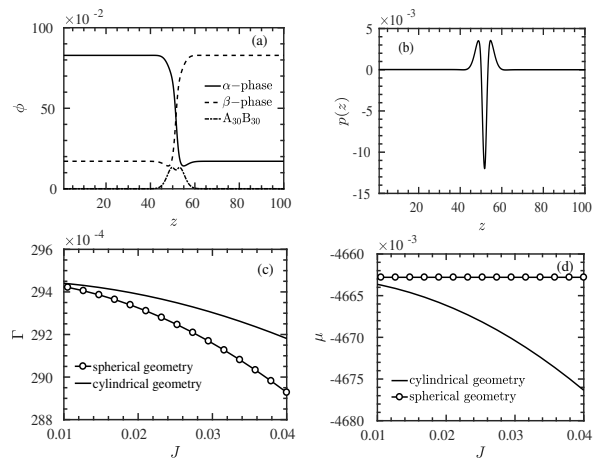}
 \caption{(a) Volume fraction distribution and (b) Lateral pressure distribution (in $k_BT/b^3$) in a planar tensionless interface. (c) Area per surfactant molecule ($\Gamma$) [in units $b^2$] 
and (d) Chemical potential of surfactant ($\mu \equiv \mu_s$) [in units of $k_BT$] as a function of interface curvature as indicated for systems with $\Delta P_L=0$. Surfactant A$_{30}$B$_{30}$, solvents, A$_{4}$ and B$_{4}$. $\chi=0.6$} 
 \label{fgr:1}
\end{figure}

Volume fraction profiles, $\varphi(z)$, and the lateral pressure distribution, $p(z)=-\omega(z)$, with  $\omega$ as the grand potential density, are presented for the default planar tensionless L/L interface, in Figs. (\ref{fgr:1}a) and (\ref{fgr:1}b). Here, $z \equiv z/b$ is the dimensionless normal coordinate. In Fig. (\ref{fgr:1}a) we see that the two liquids give a Van der Waals-like profile and the accumulated copolymers have their blocks on corresponding sides of the interface. The pressure profile $p(z)$, see Fig. (\ref{fgr:1}b), has a negative excursion at the interface due to the contribution from the L/L interface and positive `wings' on either side of the interface due to the overlap of copolymers in a brush-like configuration. From earlier work \cite{amartinus1999thermodynamic}, we know that $\gamma = -\sum_z p(z)$ and that the second moment of the pressure distribution with respect to the Gibbs plane ($R^g$) provides a direct estimate of $\bar {\kappa}=\sum -(z-R^g)^2 p(z)$. The latter relation proved useful for the evaluating $\bar {\kappa}$ of lipid bilayers, and presents a strong test for alternative, more elaborate routes to obtain the same quantity.

\begin{figure}
 \centering
 \includegraphics[width=0.375\textwidth]{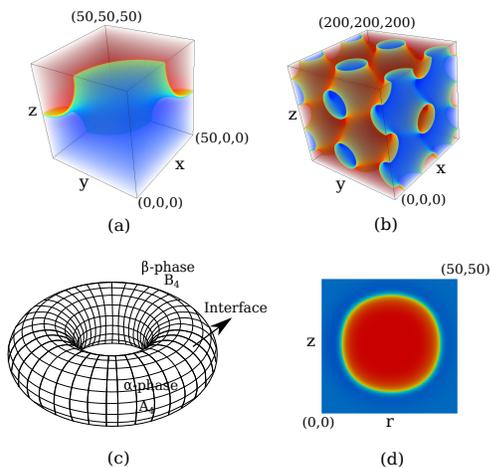}
 \caption{Volume fraction distribution of $\alpha-$phase from 3D SCF calculation of interface modeled as Im3m cubic phase. $1/8$ of a unit cell is shown in (a) 8 unit cells are shown for visualization in (b). Schematic illustration of an interface in torus shape is shown in (c). Volume fraction distribution of $\alpha-$phase from 2D SCF calculation of minimal torus in a cylindrical lattice is shown in (d). The molecular model is similar as in Fig. \ref{fgr:1}. Color scale from blue to red is $0.2-0.8$ for all contours.}
 \label{fgr:2}
\end{figure}

Similar as for lipid bilayers, an independent alternative route for evaluating $\bar{\kappa}$ makes use of three-gradient SCF computation as shown in Fig. (\ref{fgr:2}a), where on all six faces of the elementary box, Neumann boundary conditions apply; the elementary box is 1/8th of a unit cell of an Im3m phase, 8 unit cells are shown in Fig. (\ref{fgr:2}b). When equal amounts of A and B are present in the system, the interface splits the volume into two identical sub-volumes (phase $\alpha$ and $\beta$) while $J=0$ along the surface and $\Delta P_L=0$. As soon as the copolymers are added, such that the chemical potential of all molecular species is equal to the corresponding values of the planar tensionless system, we lie within an $(\mu, V, T)$-ensemble; thus, the grand potential,  $\Omega = F -\sum_j \mu_j n_j$, is the characteristic function, and $\Omega = \bar{\kappa} \int_M K dA$. Using Gauss-Bonnet theorem for compact, boundary-less Riemann manifold, the integral of curvature over the area can be evaluated as $-8\pi$ \cite{chern1944simple,fenchel1940total}. Thus, the grand potential for the unit cell, $\Omega = - 8\pi \bar{\kappa}$. Hence, from the scale-invariant grand potential $\Omega$ directly follows $\bar{\kappa}$. The result is consistent with the second moment over the pressure profile (see Tab. (1) in supplemental material \cite{supplemental}).

The procedures to evaluate $\kappa$ are more involved. In Figs. (\ref{fgr:1}c) and (\ref{fgr:1}d), we have presented typical results for spherically and cylindrically curved interfaces when $\Delta P_L =0$ as a result of the adsorption of the copolymers. In Fig. (\ref{fgr:1}c) we show the area per copolymer at the interface (inverse of the adsorbed amount) and in Fig. (\ref{fgr:1}d) the corresponding chemical potentials as a function of the curvature $J$.   
In Fig. (\ref{fgr:1}d) we notice that the chemical potentials remain constant upon bending in case of spherical curvature. This means that in this geometry bending is performed in the $(\mu, V, T)$-ensemble. The reason why the system can maintain its chemical potentials upon bending of the interface is traced to the known fact that integrating Eqn. (\ref{helfrichsfree}) over the area, $\Omega = \int_M \gamma(J,K) dA =  4\pi (2 \kappa + \bar{\kappa})$, is a constant irrespective of the size of the spherical droplet showing scale invariance. 

Now an indirect route is available to compute $\kappa$, namely from combining the total curvature energy from the spherical droplet with the Gaussian bending modulus $\bar{\kappa}$ found above. Ideally, we would like to validate this indirect route with a direct estimate.

Again, as in the cylindrical geometry, neither the adsorbed amount of surfactant, cf. Fig. (\ref{fgr:1}c), nor the corresponding chemical potential, cf. Fig. (\ref{fgr:1}d), is conserved, and we cannot use this geometry to obtain $\kappa$. A direct route to evaluate the mean bending modulus is still possible using a system that features a minimal torus, as illustrated in Fig. (\ref{fgr:2}c). Within SF-SCF this is realized using a two-gradient ($r,z$) cylindrical lattice. A typical result is presented in Fig. (\ref{fgr:2}d) as a density contour plot in the $(r,z)$ cross-section. From Gauss-Bonnet theorem, as the torus has genus $g = 1$ the integral $\int_M K dA $ vanishes. Moreover, the so-called Willmore energy of the torus has contribution only from mean curvature, $W = \frac{1}{4} \int_M J^2 dA$.

In 1965, T.J. Willmore conjectured that the Willmore energy ($W$) of a smooth torus immersed in 3D space is always greater than or equal to $2\pi^2$ \cite{willmore1965note}. This conjecture was proved by Marques and Neves in 2012 \cite{marques2012min}. The Willmore energy reaches its minimum when the radius of revolution is $\sqrt{2}$ times the radius of the generating circle, as shown in Fig. (\ref{fgr:2}d). By integrating the Helfrich equation for the toroidal configuration with minimal Willmore energy, we obtain the grand potential for torus as, $\Omega_t = \frac{1}{2} \kappa \int_M J^2 dA = 2\kappa W = 4\pi^2 \kappa$.

Now the protocol boils down to generating this minimal torus in SF-SCF while adding the copolymer such that $\Delta P_L=0$. It occurs that in this case, the system converges with all its chemical potentials equal to that of the planar tensionless interface and lies within the ($\mu, V, T$)-ensemble. Similar as in the droplet case this result is traced to the scale invariance, in this case of the minimal Willmore energy. Its grand potential gives a direct estimate of $\kappa={\Omega_t}/{4\pi^2}$.

The values found for $\kappa$ by the direct and indirect routes are congruent, proving that there is complete consistency in obtaining the bending rigidities, using scale-invariant surfaces, for tensionless balanced L/L interfaces \cite{supplemental}. Our protocol is available at \href{https://wp.me/p7KmNt-9C}{https://wp.me/p7KmNt-9C} as a open source software package.

\begin{figure}[t]
 \centering
 \includegraphics[width=0.5\textwidth]{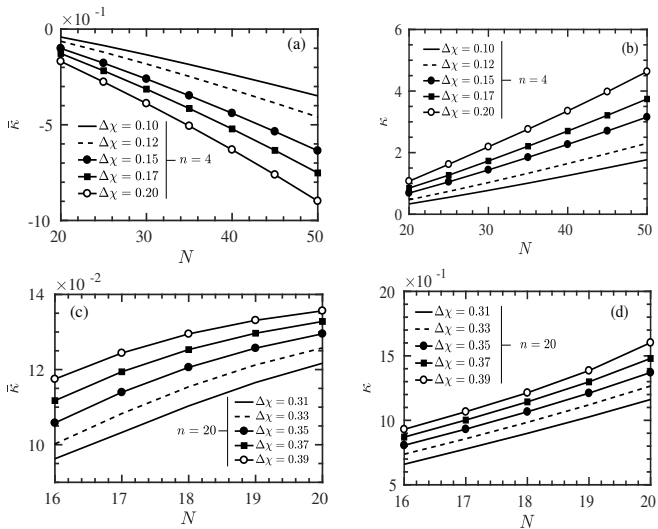}
 \caption{Chain length dependence of bending rigidities [in units of $k_BT$]. (a) and (b) $ N > n$ Regime: Chain length of bulk phases fixed (A$_4$, B$_4$), surfactant chain length is varied (A$_N$B$_N$, where $20 < N < 50$; $0.1 < \Delta\chi< 0.2$). (c) and (d) $ N \approx n$ Regime: Chain length of bulk phases fixed (A$_{20}$, B$_{20}$), surfactant chain length is varied (A$_N$B$_N$, where $16 < N < 20$; $0.3 < \Delta\chi < 0.4$).}
 \label{fgr:cd}
\end{figure}

As we have established the molecular link for bending rigidities, we now present the chain length dependence of the bending rigidities for the regime where $N > n$ in in Fig. (\ref{fgr:cd}a) and (\ref{fgr:cd}b), and for $N \approx n$  in Fig. (\ref{fgr:cd}c) and (\ref{fgr:cd}d). The trends for $N>n$ support the results from simulations \cite{PhysRevLett.92.116101,thakkar2011verifying,smit1990computer}. While it would be alluring to conclude that bending rigidities have a linear dependence on the chain length of the surfactants, results in regime $n \approx N$ contradict this observation and the dependences are clearly nonlinear, see Figs. (\ref{fgr:cd}c) and (\ref{fgr:cd}d). It is observed that the dependence of rigidities on surfactant chain length is strongly influenced by solvent chain length and interaction parameter between monomeric units, an important effect which has not been addressed in previous works \cite{PhysRevE.57.6825,kurtisovski2007molecular,szleifer1988curvature,wurger2000bending,gradzielski1996experimental,gradzielski1997droplet,safinya1989universality,PhysRevLett.92.116101,thakkar2011verifying}. A thorough analysis of the magnitude of $\kappa$ is, however, beyond the scope of the present letter and will be presented elsewhere.

Moving from the regime where the solvent length is smaller compared to the surfactant block length, $n < N$, to the regime where the solvent length is comparable to that of the surfactant, we observe that $\bar{\kappa}$ is of opposite sign, cf. Figs. (\ref{fgr:cd}a) and Fig. (\ref{fgr:cd}c). Such a sign switch is of exceptional interest, as it addresses a topological phase transition in microemulsions that can be achieved in two ways: (1) by tuning the interaction parameter for fixed solvent and surfactant lengths and (2) by tuning solvent length for fixed surfactant length and $\Delta\chi = \chi - 2/n$. 

\begin{figure}[t]
 \centering
 \includegraphics[width=0.47\textwidth]{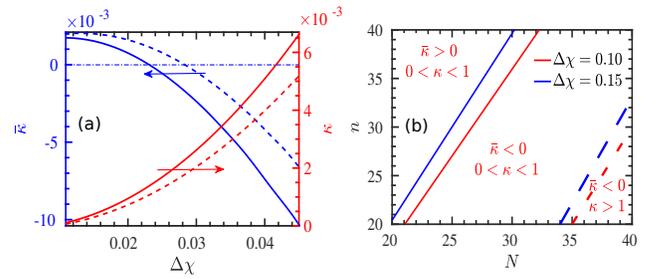}
 \caption{(a) Gaussian bending modulus (blue axis) $\bar{\kappa}$ [in units of $k_BT$] and mean bending modulus (red axis) $\kappa$ [in units of $k_BT$] as a function of $\Delta\chi$ ($\Delta\chi = \chi -2/n$). Surfactants are modeled as $A_{N}B_{N}$. $\alpha-$phase is modeled as $A_{n}$ and $\beta-$phase is modeled as $B_{n}$. [solid line: $n=4$, dashed line: $n=6$] (b) Phase diagram in $n$ and $N$ coordinates for fixed $\Delta\chi$ as indicated. The sign and magnitude [in units of $k_BT$] of the rigidities are indicated. Note that within the mean field model  $n$, $N$ are related to the radii of gyration in the bulk as $R_g =b \sqrt{n/6}$) and $R_G = b\sqrt{2N/6}$, for the solvent and copolymer respectively, where $b$ is bond length.}
 \label{fgr:3}
\end{figure}

In Fig. (\ref{fgr:3}a) the dependence of both $\kappa$ and $\bar{\kappa}$ are shown for surfactant block length of $N = 20$ as a function of a measure of closeness to the critical point of the binary solvent $\Delta\chi$. The Gaussian bending modulus, $\bar{\kappa}$, switches from negative to positive when moved towards weak segregation; this transition occurs earlier in higher $n$ (dashed line) for fixed $N$.

A similar effect can also be achieved by tuning the surfactant chain length for a given solvent chain length and interaction energy ($\Delta\chi$). Experimentally, one can reach weak segregation by the addition of a suitable co-solvent which diminishes the difference between the two primary solvents.

A summary of results, obtained by tuning $N$, is presented as a `phase diagram' in Fig. (\ref{fgr:3}b). The two governing parameters, i.e., block length of the copolymer and the chain length of the solvent are on the $x-$ and $y-$axis, respectively. The interaction parameter is chosen as $\Delta \chi$.
By tuning the surfactant length for given interactions ($\Delta\chi$) and solvent chain length, $\kappa$ increases and $\bar{\kappa}$ decreases monotonically, also showing a sign switch at the solid lines. 

These results imply the tendency of the interface to remain planar on average when $n << N$. This result is contrasted with the situation when the length of the solvent molecules is increased to be similar, $n \approx N$, or even larger than that of the copolymer, $n > N$; $\bar{\kappa}$ becomes positive in this regime while $\kappa$ is small but positive. For large $n$, we have $\bar{\kappa} >0;\ 0 < \kappa < 1$. These features are consistent with a sponge phase (Winsor III) \cite{gompper1995self} which grows in importance with reducing $\chi$. Stable but very flexible, and strongly fluctuating lamellar phases (as $\bar{\kappa}<0$ and $0<\kappa<1$) are observed as $N$ is increased for fixed $n$, whereas for very large $N$, $\kappa$ is $>1$, and we enter a region where the fluctuations of the interface are weak, crossing the dashed lines, as shown in Fig. (\ref{fgr:3}b). 

We have linked molecular characteristics to bending rigidities for surfactant-covered L/L interfaces. Using surfaces with scale-invariant curvature energies is embellished as an elegant route to determine $\kappa$ and $\bar{\kappa}$ unambiguously; this route cautiously exploits the tensionless state of the interfaces and avoids the linear term in curvature. Large deviations from these constraints imply the loss of the microemulsion middle-phase in favor of emulsions with oil-in-water or water-in-oil droplets; however, understanding the effects of small deviations is vital, as it is a prerequisite for any detailed comparison with experiments. The current analysis provides a natural starting/reference point to generalize for molecular asymmetry, spontaneous curvature and finite tension of the interfaces. 
\\

This work is part of an Industrial Partnership Programme, `Shell/NWO Computational Sciences for Energy Research (CSER-16)', of the Foundation for Fundamental Research on Matter (FOM), which is part of the Netherlands Organisation for Scientific Research (NWO). Project number: 15CSER26. We are indebted to T. E. Kodger and J. M. Clough for proofreading our letter.

\bibliography{apssamp}
\end{document}


\preprint{}
\title{Supplemental material to ``Sign switch of Gaussian bending modulus for microemulsions; a self-consistent field analysis exploring scale invariant curvature energies"}
\author{Ramanathan Varadharajan}
\email[Author to whom correspondence should be addressed: R. Varadharajan: ]{ramanathan.varadharajan@wur.nl}
\author{Frans A. M. Leermakers}
\email[F.A.M Leermakers: ]{frans.leermakers@wur.nl}
\affiliation{Physical Chemistry and Soft Matter, Wageningen University \& Research Center, Stippeneng 4, 6708WE Wageningen, The Netherlands.}
\date{\today}

\maketitle

\section*{Model and method}
\subsection*{Scheutjens-Fleer Self Consistent Field model}

The free energy $F$ is found \cite{scheutjens1979statistical} from a spatial summation of the free energy density $f({\bf r})$, where $\bf r$ is a discrete (lattice) coordinate in the system: $F  = \sum_{\bf r} f({\bf r})$. In turn, the free energy density ($f$), in units of the thermal energy, is formulated in terms of volume fraction $\varphi_i({\bf r})$ and complementary segment potential $u_i({\bf r})$ profiles for segment types $i= A,\ B$:
\begin{equation}
\begin{aligned}
F = - \ln Q([u]) - \sum_{i,{\bf r}}u_i(\textbf{r})\varphi_i(\textbf{r}) + F^{\rm int}([\varphi]) \\
+ \alpha(\textbf{r}).\bigg{[}\sum_i \varphi_i(\textbf{r}) - 1 \bigg{]}
\label{eq.scff}
\end{aligned}
\end{equation}
The system partition function is found from single chain partition functions $q_j$ for molecule component $j = A_n,\ B_n,\ A_NB_N$: $Q=\Pi_j q_j^{n_j}/n_j!$ which can be computed from the (known) segment potentials [$u_i({\bf r})$], where $n_j$ is the number of molecules of type $j$ in the system. The interaction energy 
\begin{equation}
F^{\rm int}=\sum_r n_A(r)\chi \bigg{[} \varphi_B(r) + \frac{1}{6} \nabla^2 \varphi_B(r) \bigg{]},
\end{equation}
with $n_A$ is the number of segments of type $A$. Lagrange multipliers [$\alpha(\textbf{r})$] are used to guarantee the system incompressibility at each specified coordinate. SCF solutions now involves optimizing the free energy ($F$) with respect to its variables, respectively segment potentials, volume fractions and Lagrange field.
\begin{equation}
\frac{\partial F}{\partial \varphi_i(\textbf{r})} = -u_i(\textbf{r}) + \frac{\partial F^{\rm int }}{\partial \varphi_i(\textbf{r})} + \alpha(\textbf{r}) = 0
\label{eq.op.vf}
\end{equation}
\begin{equation}
\frac{\partial F}{\partial u_i(\textbf{r})} = -\varphi_i(\textbf{r}) - \frac{\partial \ln Q}{\partial u_i(\textbf{r})} = 0
\label{eq.op.sp}
\end{equation}
\begin{equation}
\frac{\partial F}{\partial \alpha(\textbf{r})} = \sum_i \varphi_i(\textbf{r}) - 1 = 0
\label{eq.op.lp}
\end{equation}

\begin{table}
\centering
\caption{Internal consistency of estimated $\kappa$, $\bar{\kappa}$ [in units of $k_BT$]. Results are presented for surfactant chain length 30 (A$_{30}$B$_{30}$) and bulk chain length of 4 (A$_4$, B$_4$).}
\label{tab:summary}
\begin{tabular}{ l l r r }
Geometry & Invariant & $\bar{\kappa}$ & $\kappa$   \\
\hline
\hline
 & & & \\
Planar & $\sum z^2p(z) = -\bar{\kappa} $ & $ -0.14 $ & $ $  \\
Im3m  & $\Omega_i = -8\pi \bar{\kappa} $ & $ -0.15 $ & $ $  \\
Torus & $\Omega_t = 4\pi^2 \kappa$ & $ $ & $ 0.73$  \\
Sphere & $\Omega_s = 4\pi (2\kappa + \bar{\kappa})$ & $ $ & $ 0.77 $  \\
\hline
\end{tabular}
\end{table}

In this letter the molecules are modeled as freely jointed chain (FJC) \cite{fleer1993polymers}.
In such models, angular correlation between successive bonds are ignored and folding back of chain on itself is permitted. For FJC there exist an efficient propagator formalism to obtain the volume fractions [cf. Eqn. (\ref{eq.op.sp})] (and the single chain partition function $q_i$) from the potentials \cite{fleer1993polymers}.

SCF can be visualized as finding a saddle point in the multidimensional-space of segment potentials and volume fractions while ensuring incompressibility. Numerical solutions that obey to Eqns. (\ref{eq.op.vf} $-$ \ref{eq.op.lp}), have the property that the potentials both determine and follow from the volume fractions profiles and similarly the volume fractions determine and follow from segment potentials, are known to be self-consistent. Besides density distributions, we have an accurate value for the free energy. From this, all required thermodynamic properties of the system such as grand potential density 
\begin{equation}
\begin{aligned}
\omega(r) = -\sum_i \frac{\varphi_i(\textbf{r}-\varphi_i^b)}{N_i} - \alpha(\textbf{r}) -\frac{1}{2}\sum_x\sum_y \chi_{xy} \\
\times \bigg{[}\varphi_x(\textbf{r})\big{[}\varphi_y(\textbf{r}) + \frac{1}{6} \nabla^2 \varphi_y(\textbf{r})\big{]}-\varphi_x^b \varphi_y^b\bigg{]},
\end{aligned}
\end{equation}
grand potential $\Omega=\sum_r L(r) \omega(r)$ -with $L(r)$ is the number of sites at coordinate $r$- and chemical potentials 
\begin{equation}
\begin{aligned}
\mu_i = \ln \varphi_i^b + 1-N_i\sum_j \frac{\varphi_j^b}{N_j} -\frac{N_i}{2}\sum_x\sum_y \chi_{xy} \\
\times \bigg{[}(\varphi_x^b - \frac{N_{x_i}}{N_i})(\varphi_y^b - \frac{N_{y_i}}{N_i})\bigg{]},
\end{aligned}
\end{equation}
 of all species are available \cite{fleer1993polymers}.  These thermodynamical properties are used to evaluate $\kappa$ and  $\bar{\kappa}$. For detailed information of mean-field theory, refer \cite{cosgrove1987configuration,hurter1993molecular,wijmans1992self}. 

\section{Supplementary Data and results }
\subsection{Consistency of proposed protocol}
In Tab. (\ref{tab:summary}), bending rigidities obtained from different methods are exemplified for surfactant block length $N = 30$ and bulk chain length $n = 4$. It could be observed that results are internally consistent and lie within the error expected from lattice artifacts. Directly evaluated value for Gaussian bending modulus from Im3m cubic phase is consistent with estimate from planar interface, similarly the result from minimal torus is consistent with the indirect estimate of mean bending modulus from the spherical geometry.    



\bibliography{supplemental}